# Three-dimensional vortex structures and dynamics in hexagonal manganites

Fei Xue,[1] Nan Wang,[1] Xueyun Wang,[2] Yanzhou Ji,[1] Sang-Wook Cheong,[2] and Long-Qing Chen[1,*]

[1]Department of Materials Science and Engineering, The Pennsylvania State University, University Park, Pennsylvania 16802, USA

[2]Rutgers Center for Emergent Materials and Department of Physics and Astronomy, Rutgers University, Piscataway, NJ 08854, USA

*Corresponding author: lqc3@psu.edu

Hexagonal manganites $REMnO_3$ (RE, rare earths) have attracted significant attention due to their potential applications as multiferroic materials and the intriguing physics associated with the topological defects. The two-dimensional (2D) and 3D domain and vortex structure evolution of $REMnO_3$ is predicted using the phase-field method based on a thermodynamic potential constructed from first-principles calculations. In 3D spaces, vortex lines show three types of topological changes, i.e. shrinking, coalescence, and splitting, with the latter two caused by the interaction and exchange of vortex loops. Compared to the coarsening rate of the isotropic XY model, the six-fold degeneracy gives rise to negligible differences with the vortex-antivortex annihilation controlling the scaling dynamics, whereas the anisotropy of interfacial energy results in a deviation. The temporal evolution of domain and vortex structures serves as a platform to fully explore the mesoscale mechanisms for the 0-D and 1-D topological defects.

Multiferroics, with the coexistence of multiple ferroic orders, have attracted enormous attentions due to the rich physics and potential applications such as magnetoelectric random-access memory[1,2]. Recently, hexagonal REMnO$_3$ (RE, rare earths) demonstrate their intriguing multiferroic behaviors with the mutually interlocked ferroelectric, structural trimerization, and antiferromagnetic domain walls[3-5]. In contrast to conventional ferroelectrics such as BaTiO$_3$, where the Ti *3d*-O *2p* hybridization is the driving force of ferroelectricity, the polarization in REMnO$_3$ is induced by the structural trimerization during the transition from a paraelectric state (*P6$_3$/mmc*) to a ferroelectric state (*P6$_3$cm*)[6-9]. The structural trimerization gives rise to three translational phase variants based on the choice of the origin, and each variant has two options of polarization, i.e. either along +*c* or –*c* directions[10,11]. Thus, there are totally six energy-degenerate domains in the REMnO$_3$ systems. The coexistence of the six domains leads to the formation of one-dimensional (1D) and 0D topological defects, i.e. vortex lines in 3D spaces and vortices/antivortices in 2D spaces (the vortices and antivortices are categorized based on the cycling sequence of the six domains around the cores)[3,4,12].

Topological defects are solutions of partial differential equations that are topologically distinct from the uniform or trivial solution[13,14]. A topological defect is insensitive to small perturbations, and cannot be de-tangled, i.e. no continuous transformations can change a topological defect to the uniform solution. However, two topological defects tangled in opposite ways can annihilate together, in an analogy to the situation of electrons and positrons. At high temperatures, the vortex cores in REMnO$_3$ are mobile, and the vortices and anti-vortices annihilate to reduce the total interfacial energy[15-17]. Although REMnO$_3$ have been experimentally studied extensively[3,12,18,19], there are few theoretical investigations on the details of the temporal evolution of vortex-antivortex pairs and vortex lines.

In this letter, the domain structures and dynamics in REMnO$_3$ are investigated with the phase-field method, which is a powerful numerical simulation tool for the temporal evolution of microstructures[20,21]. The phase-field simulations show that the topological defects in REMnO$_3$ are real vortices, with six domains converging to one point. The evolution of the structural order parameters is demonstrated during the annihilation of vortices and anti-vortices. The vortex line evolution in 3D simulations is found to involve with three basic topological changes, i.e., shrinking, coalescence, and splitting of vortex loops. It is shown that vortex-antivortex annihilation rather than domain wall motion controls the scaling dynamics in 2D. The domain coarsening rate in 2D simulations agrees well with the prediction of the classical XY model, whereas a deviation from the isotropic XY model is found in 3D simulations. Detailed analysis demonstrates that the anisotropy of the hexagonal system is responsible for the deviation.

Hexagonal REMnO$_3$ have similar phase transitions and the same topological defects with different elements in the RE sites[15,17], and here YMnO$_3$ is studied as an example. In YMnO$_3$, the structural trimerization, as the primary order parameter, is caused by the in-plane displacements of related oxygen atoms, and can be described by the magnitude $Q$ and azimuthal angle $\Phi$ [5]. In the phase-field simulations, we transform the polar coordinates $Q$ and $\Phi$ into Cartesian coordinates $(Q_x, Q_y)$, with $Q_x = Q\cos\Phi$, $Q_y = Q\sin\Phi$. As an improper ferroelectrics, YMnO$_3$ also exhibits an induced polarization $P_z$. According to the hexagonal symmetry, the total free energy density is expressed by[5,22,23]

$$f = \frac{a}{2}(Q_x^2 + Q_y^2) + \frac{b}{4}(Q_x^2 + Q_y^2)^2 + \frac{c}{6}(Q_x^2 + Q_y^2)^3 + \frac{c'}{6}(Q_x^6 - 15Q_x^4 Q_y^2 + 15Q_x^2 Q_y^4 - Q_y^6)$$

$$- g(Q_x^3 - 3Q_x Q_y^2)P_z + \frac{g'}{2}(Q_x^2 + Q_y^2)P_z^2 + \frac{a_P}{2}P_z^2 + \frac{s_Q^x}{2}[(\frac{\partial Q_x}{\partial x})^2 + (\frac{\partial Q_x}{\partial y})^2 + (\frac{\partial Q_y}{\partial x})^2 + (\frac{\partial Q_y}{\partial y})^2] \quad (1)$$

$$+ \frac{s_Q^z}{2}[(\frac{\partial Q_x}{\partial z})^2 + (\frac{\partial Q_y}{\partial z})^2] + \frac{s_P^z}{2}(\frac{\partial P_z}{\partial z})^2 + \frac{s_P^x}{2}[(\frac{\partial P_z}{\partial x})^2 + (\frac{\partial P_z}{\partial y})^2] - E_z P_z - \frac{1}{2}\varepsilon_0 \kappa_b E_z E_z,$$

where $a, b, c, c', g, g'$, and $a_P$ are coefficients for the Landau free energy function, $s_Q^x, s_Q^z, s_P^x$, and $s_P^z$ are coefficients for the gradient energy terms, $\varepsilon_0$ is the vacuum permittivity, $\kappa_b$ is the background dielectric constant[24,25], and $E_z$ is the electric field calculated by $E_z = -\frac{\partial \varphi}{\partial z}$ with $\varphi$ the electrostatic potential. The coefficients for the Landau free energy and gradient energy are obtained from first-principles calculations[5], except that $s_P^x$ is changed from -8.88 $eV$ to 8.88 $eV$ for the efficiency of numerical calculations. Since polarization is a secondary order parameter, the modification will not change the results in the paper as validated in Supplementary Fig. S1. The background dielectric constant $\kappa_b$ is assumed to take the typical value of 50[26].

The phase-field simulations produce domain structures in the basal plane (the *xy* plane) and 3D spaces as plotted in Figs. 1(a) and 1(b), in good agreement with those observed in experimental measurements[4,12,16]. As shown in Fig. 1(a), six domains meet together at the vortex and anti-vortex cores, which are the topological defects[3]. The 3D domain structures in Fig. 1(b) exhibit an anisotropic property, i.e. the *xy* plane shows more vortex cores than the *xz* and *yz* planes. The anisotropy is specific to the hexagonal system, whose property within the basal plane is typically different from that along the *z* axis, and the anisotropy is reflected by the anisotropic gradient energy coefficients in equation (1) with $s_Q^z > s_Q^x$ [5]. Furthermore, since the polarization is along the *z* axis, a domain wall perpendicular to the *z* axis has a head-to-head or tail-to-tail configuration, which is energetically unfavorable. Thus the domain structures in Fig. 1(b) can reduce both the interfacial energy and electrostatic energy, by avoiding domain walls perpendicular to the $z$ axis. In 3D spaces, the topological defects form 1D vortex lines, and Fig. 1(c) shows the vortex lines corresponding to the domain structures in Fig. 1(b). Some vortex lines form loops insides the sample, whereas others go through the sample following the periodic boundary conditions[17,27]. The vortex lines in Fig. 1(c) tend to be parallel to the $z$ direction, which is consistent with the anisotropic domain structures in Fig. 1(b)[28,29].

With the gradient energy coefficients obtained from first-principles calculations, the phase-field models can be employed to estimate the sizes of the vortex cores by using smaller grid spacing. As shown in Figs. 2(a) and 2(b), six domains converge to a point in the basal plane, with a decreasing magnitude of the structural order parameter, which is consistent with the claim of a real vortex in earlier experimental reports[12,30]. The simulation here shows that the vortex cores have a diameter of ~ 0.49 nm (see Supplementary Fig. S2), smaller than the experimental values of 4 and 10 unit cells[12,30], which may be caused by other structural and chemical defects concentrated near the vortex cores in experimental samples. Furthermore, the simulations indicate that the vortex cores are in the high-symmetry paraelectric phase with $Q \sim 0$ and $P \sim 0$, and the physical implication is demonstrated in Figs. 2 (c) and 2(d), which are also discussed based on first-principles calculations[31]. The core structures of the structural topological defects, with reduced magnitude of $Q$, are different from those of spin vortices, e.g. magnetic skyrmions, where the spins maintain the magnitude as a constant[32].

At high temperatures, pairs of vortices and anti-vortices annihilate to reduce the total interfacial energy, and the temporal evolution can be demonstrated by the phase-field simulations. As shown in Figs. 3(a) and 3(b), firstly the vortex and anti-vortex cores approach to each other, and the magnitude of the structural order parameter between the cores is decreased. The vortex and anti-vortex cores then coalesce to one point, as illustrated in Fig. 3(c), which is in the critical state, i.e. the transient state between two topological defects and zero topological defects. The point is no longer topologically protected, since the six domains around it belong to four types of domains variants, rather than six types of domains. Afterwards, the shape of deep valley disappears and striped domains are formed, as shown in Fig. 3(d). The annihilation process from phase-field simulations is confirmed by experimental measurements, as demonstrated in Supplementary Fig. S3.

In 3D spaces, the annihilation process gives rise to the evolution of the vortex loops[33]. One classical model that shows topological defects is the XY model with 2D unit-length vectors, which are energy-degenerate with the vector pointing to any direction within the 2D plane[13]. The 2D vectors can be in 2D and 3D spaces, which correspond to 2D and 3D XY models, respectively. For simplicity, the isotropic 3D XY model is used to demonstrate the evolution of 3D vortex loops, i.e. $P_i$ is set to be 0 and $s_Q$ is set to be isotropic. Note that the system modifying will not affect the topological changes if we assume the system is infinitely large as considered by the periodic boundary conditions (see Methods).

There exist three basic topological changes, i.e. shrinking, coalescence, and splitting of vortex loops. As shown in Figs. 4(a)-4(c), a circular loop tends to shrink to reduce the interfacial energy. On the other hand, two loop segments with different centers may interact with each other, and an exchange between the two loop segments occurs. If the two loop segments are initially belong to different loops, the exchange leads to coalescence of the two loops, as shown in Figs. 4(d)-4(f). If the two loop segments are initially parts of one loop, the exchange results in the splitting of the loop, as shown in Figs. 4(g)-4(i). During the three topological changes, the numbers of vortex loops change from 1 to 0, 2 to 1, and 1 to 2, respectively, i.e., the numbers of vortex loops are increased or decreased by 1. The animation of vortex lines is demonstrated in Supplementary Moives 1-3, and the evolution of corresponding domains is provided in Supplementary Movies 4-6. As shown in Supplementary Movies 4-6, the shrinking of vortex loops is caused by the shrinking of domains, whereas the coalescence and splitting of vortex loops are accompanied by the coalescence and splitting of related domains.

The evolution of vortex loops starting from small random noises with the coefficients of YMnO$_3$ is shown in Supplementary Movie 7, which is constituted of the three basic topological changes. The evolution of the vortex lines is similar to that of the 3D lattice dislocation networks, where the dislocations with opposite Burgers vectors can annihilate and dislocation loops can shrink[34,35]. This is reasonable since physically both the vortex lines and dislocations are 1D topological defects[13].

Next we focus on the scaling dynamics during the vortex-antivortex annihilation process. The scaling hypothesis claims that at late stages, the statistics of domain structures is independent of time while the characteristic length is scaled correspondingly. It is known that the isotropic 3D XY model shows a power-law ordering kinetics, i.e. the vortex density $\rho$ is a function of simulation steps $t$ as: $\rho \sim 1/t$, whereas the 2D XY model exhibits the power-law with a logarithmic correction, i.e. $\rho \ln \rho \sim 1/t$ [36,37]. This is because the 2D XY model shows slower coarsening dynamics, which origins from the scale-dependent friction constant[36,38,39].

In the 2D simulation of YMnO$_3$ on the basal plane, the average area of the domains $A$, with $A \sim 1/\rho$, is calculated as a function of simulation steps, i.e. time $t$. Also, the results from the isotropic 2D XY model are demonstrated for comparison. As shown in Fig. 5(a), the exponent of the average area *A* with respect to time *t* at late stages is about 0.87 and 0.86 for YMnO$_3$ and the XY model, respectively, quite different from the exponent of 1.00 in the 3D XY model and grain growth models[21,37]. After considering the logarithmic correction, the exponents are both 0.96, which is consistent with the analytical predictions of the 2D XY model[36,37]. This indicates that the annihilation of vortices and anti-vortices controls the coarsening process of the vortex state, and the 6-fold degeneracy does not significantly change the scaling behavior, compared to the continuous XY model. This is because the domain walls with the 6-fold degeneracy tend to reduce their curvature, and the curvature-driven process is faster than the vortex-antivortex annihilation,[36] as demonstrated in Supplementary Movie 8. As a reference, we also provide the evolution of domain structures for the XY model in Supplementary Movie 9.

The results of 3D simulations are plotted in Fig. 5(b), and a yellow line is drawn as the reference. The simulation data for YMnO$_3$ are parallel to the yellow line at small simulation steps, whereas deviates from the yellow line for large simulation steps. To reveal the origin of the deviation, we run the simulations using the isotropic gradient energy coefficients ($s_Q^z = s_Q^x$) with all the other simulation conditions unchanged, and the results are well fitted by the yellow line, as shown in Fig. 5(b). It indicates that the simulation results with isotropic gradient energy coefficients are consistent with the prediction of the isotropic 3D XY model, and the deviation of YMnO$_3$ arises from the anisotropic gradient energy coefficients ($s_Q^z > s_Q^x$). At large simulation steps, most domains are homogeneous along the z axis as shown in Fig. 2 (b), and the coarsening process becomes 2D-like with the exponent of -0.87 (a 2D system can be regarded as homogeneous along the *z* axis with $s_Q^z \to \infty$). Therefore, due to the anisotropy intrinsic to the hexagonal system, the scaling dynamics shows a crossover from 3D-like to 2D-like behaviors.

The scaling dynamics here is different from the Kibble–Zurek mechanism reported earlier[17]. The Kibble–Zurek mechanism applies to the situation of slow cooling with different cooling rates, and is based on the critical scaling analysis above the critical point[40-42]. The scaling dynamics, on the other hand, corresponds to the phase-ordering dynamics at a specific temperature, after an infinitely rapid quench from high temperatures[36,41]. The Kibble-Zurek mechanism predicts that the vortex density follows a power-law with the exponent of ~0.57 with respect to different cooling rates[17], whereas the coarsening dynamics here gives rise to a curve with the slope ranging from -0.87 to -1.00 in the log-log plot, as shown in Fig. 5(b).

In summary, the vortex structures and scaling dynamics in hexagonal manganites are investigated by the phase-field simulations with parameters obtained from first-principles calculations. It is shown that the magnitude of the order parameters approaches to zero near the vortex cores, which indicates the high-symmetry paraelectric phase. The vortex-antivortex annihilation process in the basal plane is demonstrated by the phase-field simulations. During the coarsening process in three-dimensional (3D) spaces, the vortex loops evolve through three basic topological changes, i.e. shrinking, coalescence, and splitting, and vortex loop interaction and exchange are shown to be responsible for the coalescence and splitting. 2D scaling dynamics on the basal plane reveal that the average area of the domains with respect to the simulation steps exhibits a power-law with logarithmic correction, and the scaling dynamics is controlled by the vortex-antivortex annihilation, rather than the movement of domain walls. For 3D cases, the coarsening rate deviates from the prediction of the 3D XY model, and it is demonstrated that the deviation results from the anisotropy of the hexagonal system. Our work opens a new route to explore the detailed temporal evolution of topological defects in different dimensions.

**Methods**

The dynamical system is evolved by solving the time-dependent Ginzburg-Landau (TDGL) equations $\frac{\delta P_z}{\delta t} = L_P \frac{\delta f}{\delta P_z}$, $\frac{\delta Q_x}{\delta t} = L_Q \frac{\delta f}{\delta Q_x}, \frac{\delta Q_y}{\delta t} = L_Q \frac{\delta f}{\delta Q_y}$, where $L_P$ and $L_Q$ are the kinetic coefficients related to the domain wall mobility. The TDGL equations are solved based on a semi-implicit spectral method[43], and it is assumed that $L_P = L_Q = 0.05\,arb.\,unit$ in the simulations. The gradient energy coefficients are normalized based on $s_Q^{*z} = s_Q^{*z}/g_0$, $s_Q^{*x} = s_Q^{*x}/g_0$, $s_P^{*z} = s_P^{*z}/g_0$, and $s_P^{*x} = s_P^{*x}/g_0$, where $g_0 = a(\Delta x)^2$. Thus the normalized gradient energy coefficients are dependent on the grid spacing. For the simulation of scaling dynamics, the initial conditions are small random noises for the order parameter components, which represents that the system is quenched from high temperatures. Periodic boundary conditions are applied along the three directions.

**Acknowledgement:**

The work at Penn State is supported by the NSF MRSEC under Grant No. DMR- 1420620 and DMR-1210588. The work at Penn State used the Extreme Science and Engineering Discovery Environment (XSEDE), which is supported by National Science Foundation grant number ACI-1053575[44]. The work at Rutgers is funded by the Gordon and Betty Moore Foundation's EPiQS Initiative through Grant GBMF4413 to the Rutgers Center for Emergent Materials.

**Supplementary Information** is linked to the online version of the paper at www.nature.com/nature

**Author contributions**

L.-Q. C. and S.-W. C. designed the research project and supervised the simulations and experiments. F. X. and Y. J. carried out phase-field simulations. F. X. and N. W. analyzed the simulation results. X. W. collected and analyzed experimental data. All authors discussed the results, and co-wrote the paper.

**Additional information**

The authors declare no competing financial interests. Correspondence and requests for materials should be addressed to L.-Q. C..

**Figure Legends**

**Figure 1 | Domain configurations from the phase-field simulations.** (a) Part of the domain structures from a 2D simulation on the basal plane with a system size of $1024\Delta x \times 1024\Delta x \times 1\Delta x$ and grid spacing of $\Delta x = 0.30\, nm$. The six colors represent six types of domains. $\alpha$, $\beta$, and $\gamma$ denote the three structural trimerization phases, and + and – superscripts indicate the direction of the polarization. (b) Domain structures from a 3D simulation with a system size of $128\Delta x \times 128\Delta x \times 128\Delta x$ and grid spacing of $\Delta x = 0.30\, nm$. (c) Vortex lines in 3D spaces corresponding to the domain structures in (b).

**Figure 2 | Demonstration of vortex core structures.** (a) and (b) Distribution of the structural trimerization magnitude $Q$ near a vortex core and an anti-vortex core, respectively. The domain structures are from 2D simulations on the basal plane with grid spacing of $\Delta x = 0.020\, nm$, and the height represents the value of $Q$ with the equilibrium bulk value of $9.5 \times 10^{-2}$ $nm$. (c) and (d) The corresponding atomic structures near a vortex core and an anti-vortex core, respectively. The vortex cores have a size of ~ one unit cell, and show the paraelectric phase with $Q$ ~ 0. The red arrows in (c) and (d) represent the displacement directions of oxygen atoms.

**Figure 3 | Evolution of vortex structures during vortex-antivortex annihilation.** (a)-(d) Vortex structures at different simulation time steps. The simulation is on the 2D basal plane and the height represents the magnitude of the structural order parameter $Q$. In (c) the two red domains are connected by a point with Q~0, and the point separates the two light blue domains as well.

**Figure 4 | Three basic topological changes of vortex lines in an isotropic 3D XY model.** (a)-(c) Shrinking, (d)-(f) coalescence, and (g)-(i) splitting of vortex loops.

**Figure 5 | Coarsening dynamics from the phase-field simulations.** (a) Average domain area as a function of simulation steps in the 2D basal plane. The squares and circles correspond to the raw data and the data with logarithmic correlations, respectively. Different lines are fitted based on the formula: $a + t^b$. (b) Total vortex line length as a function of simulation steps in 3D simulations. The yellow circles are results using isotropic gradient coefficients as the control experiments, which are fitted by a yellow line. The pink squares correspond to the anisotropic gradient coefficients in the hexagonal YMnO$_3$. (a) is the average result of five parallel simulations with a system size of $2048\Delta x \times 2048\Delta x \times 1\Delta x$ and grid spacing of $\Delta x = 0.30\, nm$, and (b) is the average of three parallel simulations with a system size of $512\Delta x \times 512\Delta x \times 512\Delta x$ and grid spacing of $\Delta x = 0.30\, nm$.

**Figure 1**

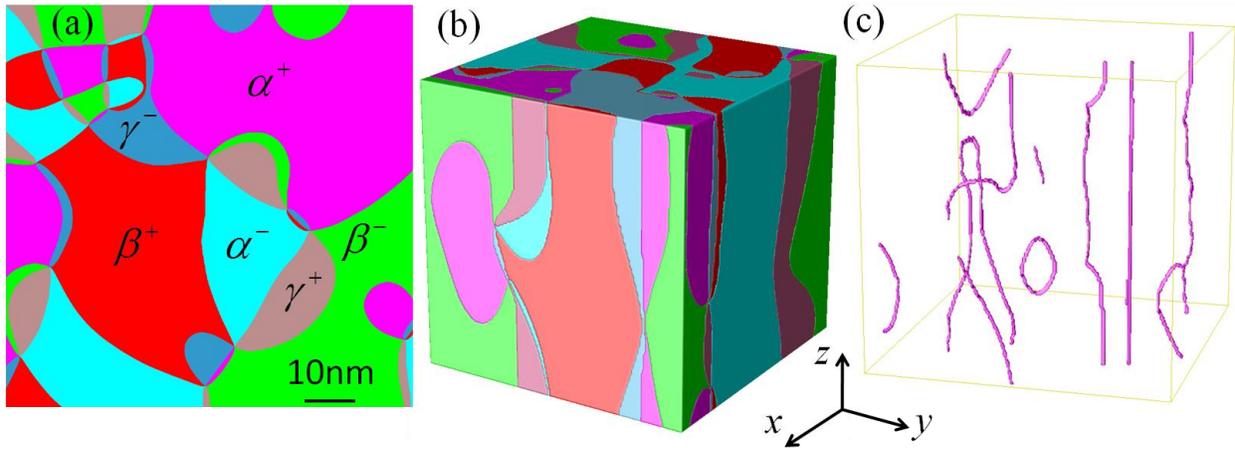

**Figure 2**

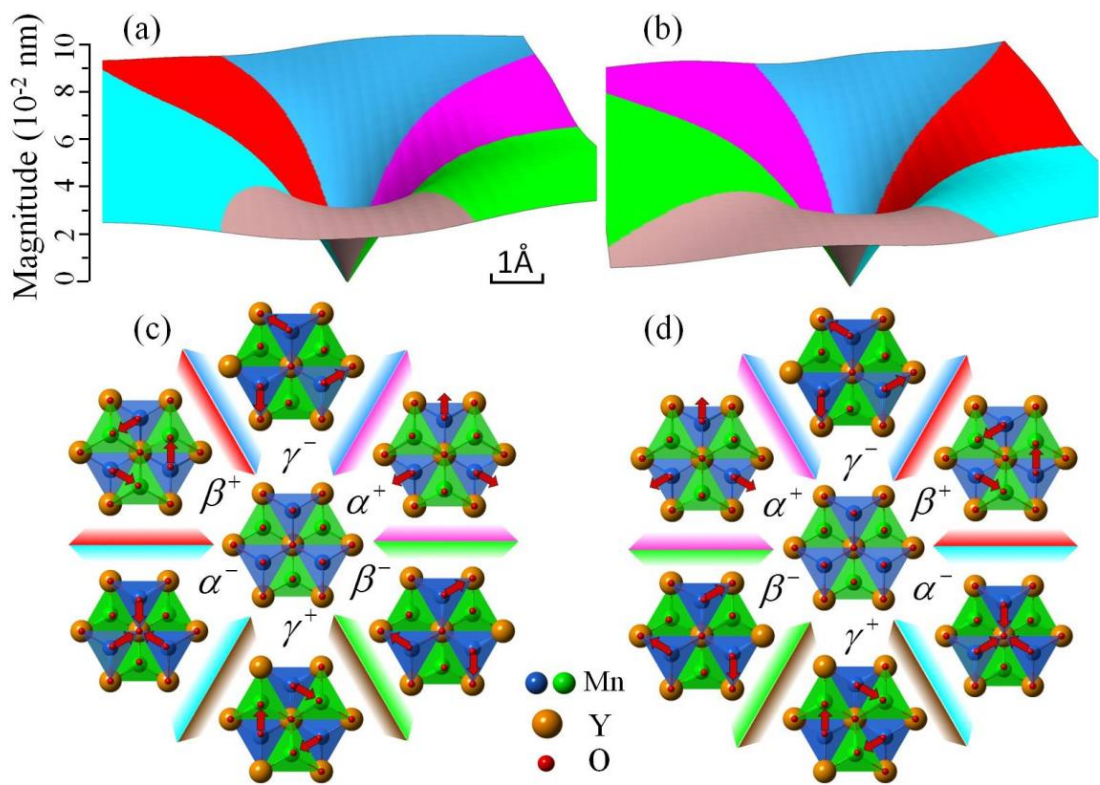

**Figure 3**

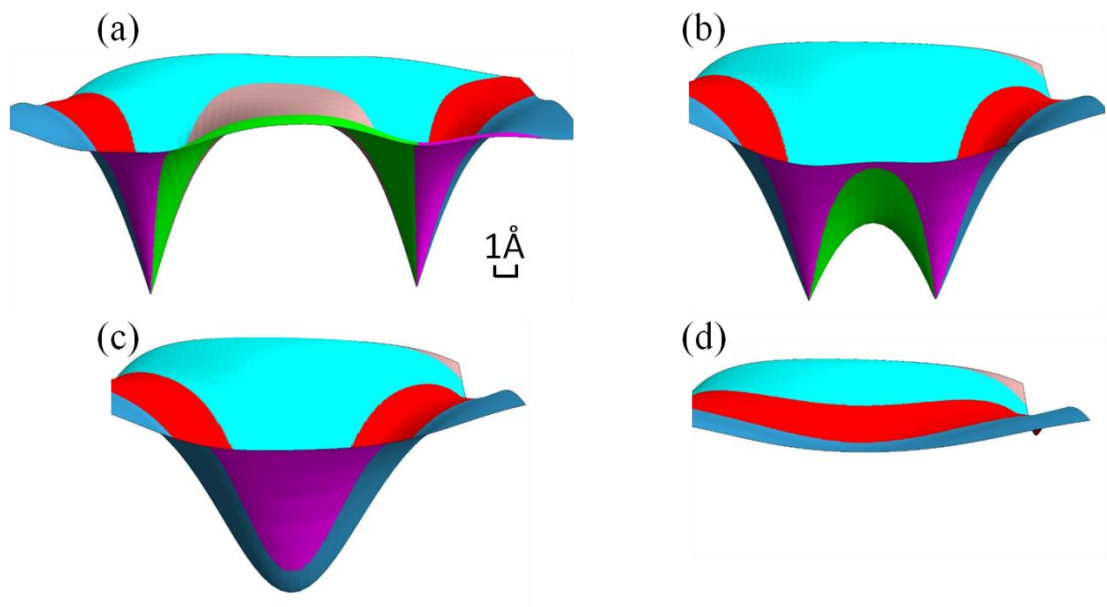

**Figure 4**

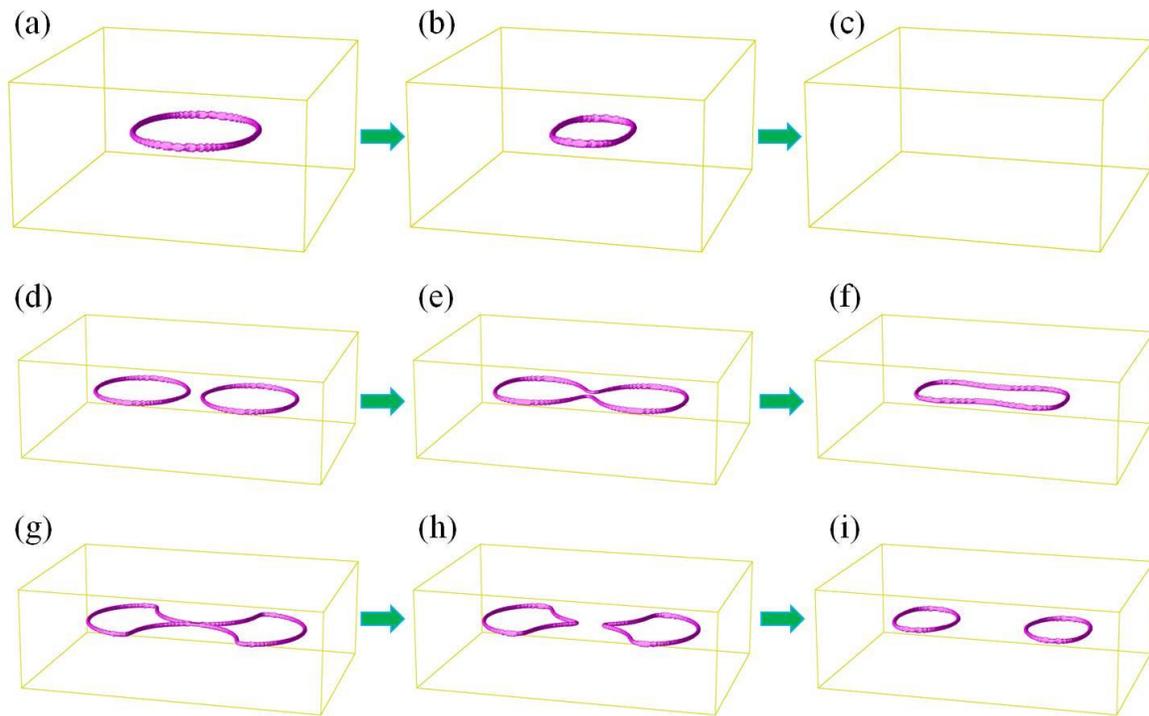

**Figure 5**

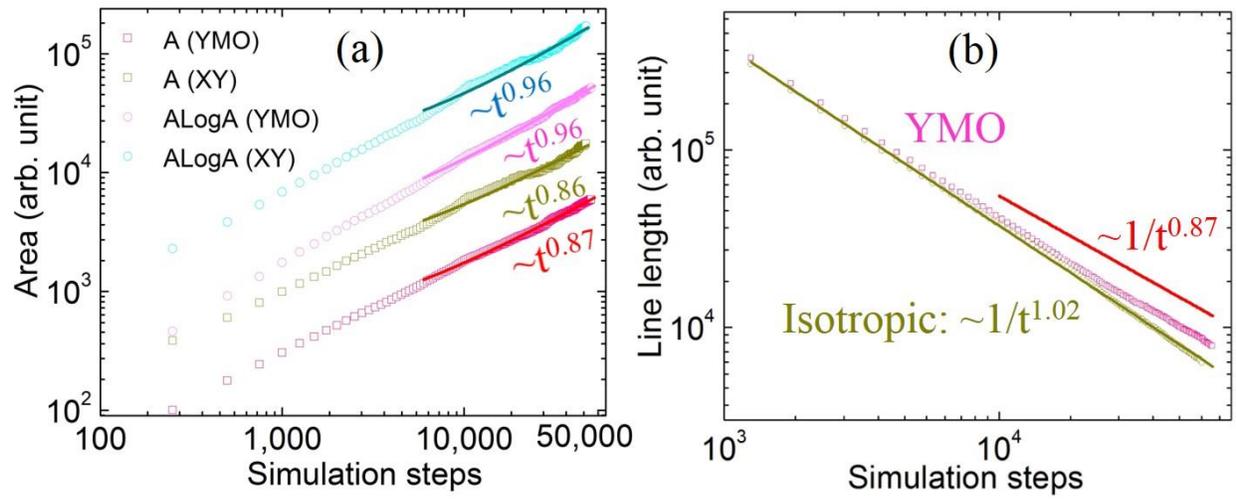